\documentstyle[epsfig,11pt]{article}

\newcommand{\fig}[1]{Fig.\ref{#1}}
\newcommand{\tab}[1]{table \ref{#1}}

\newcommand{\eqn}[1]{Eq.(\ref{#1})}
\newcommand{\gev}{\mbox{ GeV}}

\newcommand{\ben}{\begin{enumerate}}
\newcommand{\een}{\end{enumerate}}
\newcommand{\bit}{\begin{itemize}}
\newcommand{\eit}{\end{itemize}}
\newcommand{\bc}{\begin{center}}
\newcommand{\ec}{\end{center}}
\newcommand{\bb}{\begin{bf}}
\newcommand{\eb}{\end{bf}}
\newcommand{\bsm}{\begin{small}}
\newcommand{\esm}{\end{small}}
\newcommand{\bns}{\begin{normalsize}}
\newcommand{\ens}{\end{normalsize}}
\newcommand{\bq}{\begin{equation}}
\newcommand{\eq}{\end{equation}}
\newcommand{\bqa}{\begin{eqnarray}}
\newcommand{\eqa}{\end{eqnarray}}

\newcommand{\vb}{\vspace*{2cm}}

\newcommand{\nn}{\nonumber}
\newcommand{\mtt}[1]{\mbox{\tt{#1}}}

\def\awf{a_{W\Phi}}
\def\abf{a_{B\Phi}}
\def\aw{a_{W}}
\def\b{{\cal B}}

\def\ctg{\mbox{ctg~}}

\def\c2{\chi^2}
\def\SM{Standard Model\ }
\def\notp{p\hspace*{-5.5pt}/}

\def\L{{\cal L}}
\def\O{{\cal O}}

\def\onehalf{{1\over 2}}
\def\vtau{\mbox{\boldmath $\tau$}}
\def\vW{\mbox{\boldmath $W$}}
\def\kg{\kappa_\gamma}\def\kZ{\kappa_Z}
\def\xg{x_\gamma}\def\xZ{x_Z}
\def\dZ{\delta_Z}\def\lg{\lambda_\gamma}
\def\lZ{\lambda_Z}
\def\bqp{\bar{q}'}

\def\demo{$\Delta\eta\mu \acute{o} \kappa \varrho \iota \tau o \varsigma$}
\newcommand{\trsub}[1]{ {_{_{ {#1} }}} }

\begin{document}
\input feynman
 
\pagestyle{empty}
 
\begin{flushright}
DEMO-HEP-96/01 \\
{\tt hep-ph/9605303} \\
May 1996
\end{flushright}
 
\vspace*{2cm}
\bc\begin{LARGE}
{\bf
Studying trilinear gauge couplings at LEP2 \\
using optimal observables.
}\\ \end{LARGE}
\vspace*{2cm}
{\Large
Costas G.~Papadopoulos
} \\[12pt]
Institute of Nuclear Physics, NRCPS \demo, 15310 Athens, Greece
\\
and \\
CERN, Theory Division, CH-1211 Geneva 23, Switzerland\\[12pt]
\vb
ABSTRACT\\[12pt]  \ec
\begin{quote}

We study the sensitivity of the processes 
$e^+e^-\to\ell\bar{\nu}_\ell\; q \bqp$ at LEP2 energies on the
non-standard trilinear gauge couplings (TGC), using
the optimal observables method. 
All relevant leading logarithmic corrections to the tree-order
cross section, as well as experimental resolution effects
have been studied. Taking into account correlations among the
different TGC parameters we show that the limits on the TGC can reach the
level of 0.15~(1sd) at 161 GeV with 100 pb$^{-1}$, a challenge 
for the first LEP2 phase. At higher energies this can be
improved drastically, reaching the level of 0.02~(1sd).

\end{quote}
\vspace*{\fill}
\noindent DEMO-HEP-96/01 \\
\noindent May 1996
\newpage
\pagestyle{plain}
\setcounter{page}{1}

\par

One of the two most important measurements at LEP2 energies
will be the determination of the trilinear vector boson
couplings \cite{goun:0,bile:0,hagi:0,papa:sngw},
a characteristic manifestation of the underlying
non-Abelian symmetry of elementary particle interactions
\cite{sta-mod}.
 
\par
In order to study the trilinear boson couplings 
we need a parametrization of these
interactions that goes beyond the \SM. There are, of course,
several possibilities to accomplish this task, but we are going to restrict 
ourselves to the
most economical one  by considering C$-$ and P$-$ conserving interactions.
The relevant interaction Lagrangian 
is usually written in the following form \cite{bile:0}:
\bqa
\L_{TGC}&=&\sum_{V=\gamma,Z} e\: g_V\left(
V_\mu W^{-\mu\nu}W^{+}_{\nu}
-V_\mu W^{+\mu\nu}W^{-}_{\nu}
+\kappa_V V_{\mu\nu}W^{+\mu}W^{-\nu}\right)\nn\\
&+&e\:{\lambda_\gamma\over M_W^2}A_{\mu\rho}W^{+\rho\nu}W^{-\mu}_\nu
+e\:\ctg\theta_w{\lambda_Z\over M_W^2}Z_{\mu\rho}W^{+\rho\nu}W^{-\mu}_\nu
\label{lagrangian}
\eqa
where
\[ V_{\mu\nu}=\partial_\mu V_{\nu}-\partial_\nu V_{\mu},\;\;
W^\pm_{\mu\nu}=\partial_\mu W^\pm_{\nu}-\partial_\nu W^\pm_{\mu}\;\;.
\]
$W^{\pm}$ is the $W$-boson field, and
$g_\gamma=1$, $g_Z=\ctg\theta_w$, $\kg=\kZ=1$ and
$\lg=\lZ=0$ at tree order in the \SM.
It is more convenient to express the different couplings
in terms of their deviations from the \SM values. For this we
define the following deviation parameters\cite{bile:0}:
\bq
\dZ=g_Z-\ctg\theta_w\;\;\;\;
\xg=\kg-1\;\;\;\;
\xZ=(\kZ-1)(\ctg\theta_w+\dZ)\;.
\label{deviation}
\eq
It is worth while to note that the interaction Lagrangian
becomes linear with respect to the above parameters (including also
$\lg$ and $\lZ$).
\par
During the last years, considerable progress has been achieved
concerning the understanding of the physics underlying
the non-standard boson self-couplings.
As Gounaris and Renard \cite{goun:gauge} showed,
the deviations from the \SM couplings can be parametrized in a
manifestly gauge-invariant (but still non-renormalizable)
way, by considering gauge-invariant operators involving higher-dimensional
interactions among gauge bosons and Higgs field. These operators
will be scaled by an unknown parameter $\Lambda_{NP}$, which might be
understood as the characteristic scale of New Physics effects.
In order to describe all five C$-$ and P$-$ conserving couplings
introduced in \eqn{lagrangian}
we need operators with dimension up to eight. On the other hand
restricting ourselves to
$SU(2)_L\times U(1)_Y$-invariant operators with dimension up to six,
which are the lowest order ones in $1/\Lambda_{NP}$ expansion,
we can have the following list \cite{pap:sw-ww}:
\bqa
\O_{B\Phi}&=&B^{\mu\nu}(D_\mu\Phi)^\dagger(D_\nu\Phi)\nn\\
\O_{W\Phi}&=&(D_\mu\Phi)^\dagger\;\vtau\cdot\vW^{\mu\nu}
(D_\nu\Phi)\nn\\
\O_W&=&{1\over 3!}(\vW^{\mu}_{\;\rho}\times\vW^{\rho}_{\;\nu})
\cdot\vW^{\nu}_{\;\mu}
\label{operator}
\eqa
where $\tau_i=\onehalf\sigma_i$ ($\sigma_i$ are the Pauli matrices),
\bqa
B_{\mu\nu}=\partial_\mu B_{\nu}-\partial_\nu B_{\mu}
\nn\eqa
where $B_\mu$ is the $U(1)_Y$ gauge field,
\bqa
\vW_{\mu\nu}=\partial_\mu \vW_{\nu}-\partial_\nu \vW_{\mu}
-g\vW_{\mu}\times\vW_{\nu}\nn\eqa
where $\vW$ are the $SU(2)_L$ gauge fields and
\bqa
\Phi=\left( \begin{array}{c} \phi^+\\
{1\over \sqrt{2}}(v+H+i\phi^0)\end{array}\right)
\nn\eqa
is the Higgs doublet.
The covariant derivative $D_\mu$ is given, as usual, by
\bqa
D_\mu=\partial_\mu+i\; g\vtau\cdot\vW_\mu-i\; g^\prime B_\mu
\nn\eqa
and $e=g\sin\theta_w=g^\prime\cos\theta_w$.
\par
The interaction Lagrangian can be written now as
\bq
\L_{TGC}=
i\,g^\prime {\abf\over m_W^2}\O_{B\Phi}
+i\,g{\awf\over m_W^2}\O_{W\Phi}
+g{\aw\over m_W^2}\O_W
\label{gi-lagrangian}
\eq
where the relations between $\awf$, $\abf$, $\aw$ and the deviation
parameters of \eqn{deviation} are given by:
\bqa
&\dZ=\awf/\left( \sin\theta_w \cos\theta_w\right)\;\;\;\;
\xg = \abf+\awf \;\;\;\;
\lg =  \aw &           \nn\\                        
&\xZ = -\tan\theta_w \xg   \;\;\;\;
\lZ = \lg&
\label{relation}
\eqa


\par
In order to study the effect of TGC, one traditionally considered the
reaction $e^+ e^-\to W^+ W^-$, taking into account the subsequent
decay of the $W$'s in a four-fermion final state. 
These final states can be classified in three categories, namely the `leptonic' 
$\ell_1 \bar{\nu}_{\ell_1} \ell_2^* \nu_{\ell_2}$,
the `semileptonic' $\ell\bar{\nu}_\ell\; q \bqp$ and the
hadronic $ q_1 \bqp_1 \bar{q}_2 q^\prime_2$ ($q$ and $q^\prime$ refer to up- and
down-type quarks respectively). 
Semileptonic seems to be the most favoured channel \cite{lep2tgc} for 
studying TGC, since
it contains the maximum kinematical information, taking into account that
charge-flavour identification in four jet channel is rather inefficient and
the cross section for the leptonic mode is suppressed.
It is the goal of the present paper to 
study the effect of TGC, in their three parameter version \eqn{gi-lagrangian}, 
in the processes 
\bq e^+e^-\to\ell\bar{\nu}_\ell\; q \bqp 
\label{processes}
\eq
where $\ell$ is an electron or a muon, at LEP2 energies, based on the
four-fermion Monte Carlo generator ERATO\cite{pap:4f,lep2gen,erato}. The final state 
$\tau \bar{\nu}_\tau\; q \bqp$ will
not be considered here due to the special difficulties to identify
$\tau$'s in this environment.

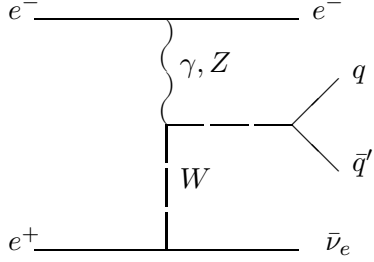
\begin{figure}
\unitlength=0.01pt
\bc
\begin{picture}(10000,8000)
\global\Xone=5000
\global\Xtwo=2500
\global\Yone=3
\global\Ytwo=4

\pfrontx = 0
\pfronty = 2000

\drawline\fermion[\E\REG](\pfrontx,\pfronty)[10000]
\global\advance \pfrontx by -1000 
\put(\pfrontx,\pfronty){$e^+$}
\global\advance \pbackx by 1000 
\put(\pbackx,\pbacky){$\bar{\nu}_e$}
\global\gaplength=250
\drawline\scalar[\N\REG](\pmidx,\pmidy)[3]
\global\advance \pmidx by 500 
\put(\pmidx,\pmidy){$W$}

\drawline\photon[\N\REG](\pbackx,\pbacky)[\Ytwo]
\global\advance \pmidx by 500 
\put(\pmidx,\pmidy){{$\gamma,Z$}}
\global\gaplength=250
\drawline\scalar[\E\REG](\pfrontx,\pfronty)[3]
\drawline\fermion[\NE\REG](\pbackx,\pbacky)[\Xtwo]
\global\advance \pbackx by 500 
\put(\pbackx,\pbacky){$q$}
\drawline\fermion[\SE\REG](\pfrontx,\pfronty)[\Xtwo]
\global\advance \pbackx by 500 
\put(\pbackx,\pbacky){$\bar{q}'$}

\drawline\fermion[\E\REG](\photonbackx,\photonbacky)[5000]
\global\advance \pbackx by 500 
\put(\pbackx,\pbacky){$e^-$}
\drawline\fermion[\W\REG](\pfrontx,\pfronty)[5000]
\global\advance \pbackx by -1000 
\put(\pbackx,\pbacky){$e^-$}

\end{picture}
\ec
\caption[.]{Single-resonant graph where TGC are contributing.}
\label{fig0}
\end{figure}

\par From the point of view of the four-fermion production, 
$e^+ e^-\to W^+ W^-$ calculations are equivalent to the narrow
width approximation, $\Gamma_W\to 0$, which requires
that both $W$'s are
kinematically allowed to be on-shell. This last requirement implies
of course that the overall energy should be above the $2 m_W$ threshold
and therefore it is not adequate for the first phase of LEP2, where 
$\sqrt{s}=161$~GeV.
Moreover in $e^+ e^-\to W^+ W^-$ the invariant masses of the produced fermion pairs, 
$\ell\bar{\nu}_\ell$ and $q \bqp$, are identical to the
$W$ mass, which requires experimental selection algorithms, whose 
efficiency is some times questionable, and more importantly their interplay 
with TGC studies can not be estimated, unless a full four-fermion 
calculation is used.
On the other hand there are nowadays widely available  
four-fermion calculations\cite{pap:4f,berends-an,lep2gen} where TGC are included beyond the 
narrow width approximation. This enable us to study TGC not only in
the double-resonant (CC3) graphs contributing to $e^+ e^-\to W^+ W^-$ but 
also in the single-resonant ones shown in \fig{fig0}, which moreover become dominant
at very high energies.  

\par 
In the actual calculations presented in this paper we have used the ERATO
Monte-Carlo generator. A detailed description of ERATO can be found
in references \cite{pap:4f,lep2gen,erato}. The basic ingredients of the
calculation are the following:
\ben
\item Exact tree-order matrix elements for the processes 
$ e^+e^-\to\ell\bar{\nu}_\ell\; q \bqp$, based on helicity amplitude calculations
\cite{pap:sw-ww,pap:4f}, including all TGC interactions described by
\eqn{lagrangian}.
\item Phase-space generation algorithm based on a multi-channel 
Monte Carlo approach, including 
weight opti\-mi\-zation\cite{weightopt}.
\item Treatment of unstable particle ($W$ and $Z$)
width consistent with gauge-invariance and high-energy 
unitarity \cite{pap:4f,baur+zep,bhf}.
\item Initial state radiation (ISR) based on the structure function
approach\cite{lep2ww,lep2gen}, including soft-photon exponentiation 
as well as leading logarithmic (LL) contributions
up to order $O(\alpha^2)$.
\item Coulomb correction\footnote{For a detailed description see 
ref.\cite{lep2ww}, pp 117-119.}
to the double resonant (CC3) graphs \cite{lep2ww}.
\een
\par Moreover in order to avoid matrix element singularities as well as to
be as close as possible to the experimental situation, 
the `canonical cuts' of reference \cite{lep2gen} have 
been applied:
\bq
175^o  \geq  \theta_\ell\;,\;\theta_{jet}  \geq 5^o
,\;\;\;
E_\ell \geq  1 \gev\;,\;\; E_{jet} \geq 3 \gev\;\;\mbox{and}\;\;
m_{q,\bqp}\geq 5 \gev
\label{cuts}
\eq
\par The input Standard Model parameters we have used are as follows:
\bqa 
&M_W = 80.23\gev,\;\;\Gamma_W=2.033\gev,\;\;
M_Z=91.188\gev,\;\;\Gamma_Z=2.4974\gev,&\nn \\
&\sin^2\theta_w = 0.23103\;\;\mbox{and}\;\; \alpha=1/128.07&
\label{inputs}
\eqa
As far as the ISR structure function is concerned the value 
$\alpha=1/137.036$ is used.


\par Most of the techniques used up to now to determine the sensitivity of
a given reaction on the TGC have been based on a Maximum-Likelihood fit to 
Monte-Carlo generated four-fermion data. In its most advanced version
\cite{sekulin}, which is 
capable of dealing with low statistics data without any binning
procedure, one maximizes the log-likelihood function:
\bq \ln \L_{ML}=\sum_{i} \ln p(\Omega_i,\vec{a})\;\;,
\eq 
where the sum is running over the event sample.
$\Omega_i$ represents the collection of the independent 
kinematical variables describing the i-th event, 
$\vec{a}=(\awf,\abf,\aw)$ is the vector whose components are 
the three TGC parameters, 
$p(\Omega_i,\vec{a})$ is the
probability to find an event at such a phase space point, and is
defined by:
\bq p(\Omega,\vec{a})={1\over \sigma}{d\sigma\over d\Omega}
\eq
with 
\[ \sigma=\int_{V} {d\sigma\over d\Omega} d\Omega\;\;.
\]
Although this is the most direct method to analyse experimental data,
one has to calculate the full differential cross section for a large
number of TGC parameter values, which is rather time-consuming.
\par 
On the other hand, since the interaction Lagrangian is linear with respect to
TGC, one can easily write the differential cross section in the form
\bq 
{d\sigma \over d\Omega}=c_0(\Omega)+\sum_{i}a_i c_{1,i}(\Omega)
+ \sum_{i,j}a_i a_j c_{2,ij}(\Omega)
\label{differentialxs}
\eq
and of course in a similar way the total cross section is
written as 
\bq 
\sigma=\hat{c}_0+\sum_{i}a_i \hat{c}_{1,i}
+ \sum_{i,j} a_i  a_j \hat{c}_{2,ij}
\eq
where hatted $c$'s are integrals of unhatted ones over the phase space.
It is now straightforward to verify that the probability has an expansion
\bq 
p(\Omega,\vec{a})=p_0(\Omega) \left\{ 1+\sum_{i}
 a_i \left({c_{1,i}(\Omega)\over c_0(\Omega)}-
{\hat{c}_{1,i}\over \hat{c}_0}\right)
+ \mbox{higher orders} \right\}
\label{probability}
\eq
where
\bq 
p_0(\Omega)={1\over \sigma}{d\sigma\over d\Omega}\bigg|_{\vec{a}=0}
\eq
is the \SM distribution.
\par 
Assuming now that the data are well described by the \SM ($\vec{a}=0$) 
distribution it is easy to calculate the so-called information matrix \cite{eadi},
which determines the sensitivity of the fit on the parameters 
$\vec{a}$, and is defined by:
\bqa 
I_{ij}&\equiv&E\bigg[ \left({\partial\over \partial  a_i} \ln\L_{ML}\right)
\left({\partial\over \partial  a_j} \ln\L_{ML}\right) \bigg] \\
&=& - E\bigg[ {\partial\over \partial  a_i}{\partial\over \partial  a_j}
\ln\L_{ML}\bigg]\nn
\label{infomatrix}
\eqa

where
\bq 
E[ A] = \int \prod_{i=1}^{N} \{ d\Omega_i\;p_0(\Omega_i)\} A(\Omega_1,...,\Omega_N)
\eq
represents the mean value of a function $A$.
To the lowest order one has that $I_{ij}=N \b_{ij}$ where
\bq 
\b_{ij}\equiv  \langle { c_{1,i}\over c_0} { c_{1,j}\over c_0} \rangle\trsub{0}
-\langle { c_{1,i}\over c_0}\rangle\trsub{0} 
 \langle { c_{1,j}\over c_0}\rangle\trsub{0}
\eq
and 
\bq
\langle A \rangle\trsub{0} = \int d{\Omega}\; p_0(\Omega)\; A(\Omega)\;\;.
\eq
Using now \eqn{probability} one has that the set of the phase-space
functions:
\bq
\O_i= { c_{1,i}(\Omega)\over c_0(\Omega)} \;\;,
\label{optimal}
\eq
are the optimal observables\cite{optobs}, whose mean values and 
covariance matrix will determine the sensitivity on the TGC parameters,
and it is equivalent, to the lowest order, to the log-likelihood method.
More specifically we have that
\[
\langle \O_i \rangle = \langle \O_i \rangle\trsub{0} +\sum_j  a_j \b_{ij}. 
\]
The estimator is now given by
\[
\bar{ a}_i=\sum_j \b^{-1}_{ij} \left( \langle \O_j \rangle 
- \langle \O_j \rangle\trsub{0}
\right)
\]
and the corresponding covariance matrix
\[
V(\bar{ a} )= {1\over N}\;\; \b^{-1} \cdot V(\O ) \cdot \b^{-1}
\]
where $V(\O )$ is the covariance matrix of the $\O_i$ defined by
\[
V(\O)_{ij} =   \langle \O_i \O_j \rangle -
 \langle \O_i\rangle \langle \O_j\rangle
\]
which in the ideal case, where measured distributions are identical to the
\SM predictions, is given by $V(\O)=\b$.
\par One-dimensional investigations assuming that all but one of the $ a$'s
are non-vanishing, result in parameter errors (1sd) given by
\bq 
\delta a_i={1\over\sqrt{N \b_{ii}}} \;\;.
\eq
On the other hand taking into account the correlations one has to
diagonalize the covariance matrix $V(\bar{ a})$ and therefore
\bq 
\delta a^D_i={1\over\sqrt{N \lambda_i}} 
\eq
where $\lambda_i$ are the eigenvalues of the matrix $\b$ and the
parameters $ a^D_i$ are defined by
\bq 
 a^D_i=\vec{e}_i\cdot\vec{a} \;\;.
\label{orthogonal} 
\eq
where $\vec{e}_i$ are the corresponding eigenvectors of $\b$.

\par In all calculations presented in this paper, $N$ is taken to be 
the predicted number of events in the \SM, defined as
\[
N=4\;L\;\sigma 
\]
where $\sigma$ is the corresponding total cross section, 
$L$ is the integrated luminosity and the factor $4$ takes into account 
the four equivalent channels (all described by the same matrix elements), 
i.e. $ e^+e^-\to\ell\bar{\nu}_\ell\; q \bqp$
stands for $ e^+e^-\to$, $\ell\bar{\nu}_\ell\; u \bar{d}$,
$\ell\bar{\nu}_\ell\; c \bar{s}$, $\ell^*{\nu}_\ell\; d \bar{u}$
and $\ell^*{\nu}_\ell\; s \bar{c}$.

\par At this point it is worth to emphasize that expanding the log-likelihood
function to higher orders in $ a_i$ will allow us to identify observables 
which probe the behaviour of $\log\L_{ML}$ up to 
the corresponding order. Therefore there is no limitation, in principle, 
to establish a one-to-one mapping between Maximum-Likelihood and 
optimal observables.
On the other hand one can argue that since we have truncated our
interaction lagrangian to the order $1/\Lambda_{NP}^2$, consistency would 
require that each observable should also be trusted up to this order and
therefore only linear terms should be kept. In any case optimal observable
and  Maximum-Likelihood become identical
in the high statistics limit,
$N\to\infty$, and therefore we are well justified to use optimal observables
to estimate the errors on TGC parameters. 
Moreover, the 
optimal observables method can easily be extended\cite{optobs}, when the variation 
of the total event rate is taken into account, corresponding to the
so-called Extended Maximum Likelihood (EML) method.
It should be noted, however, that in the case where some of 
the $c_{1,i}$ terms vanish, or 
$
\langle ({c_{1}\over c_0})^2\rangle-\langle {c_{1}\over c_0}\rangle^2 
\sim O(1/\sqrt{N})\times
( \langle {c_{2}\over c_0}{c_{1}\over c_0} \rangle-
\langle {c_{2}\over c_0}\rangle\langle {c_{1}\over c_0} \rangle ), 
$ 
leading and next-to-leading terms become equally important and an analysis based 
on the optimal observables of \eqn{optimal}
breaks down.
%
%
Fortunately, such a `pathological' case is not observed at LEP2 energies.   

\par
Finally, from the point of view of a weighted Monte-Carlo
approach, which is frequently used in the phenomenological analyses,
optimal observables offer a very efficient, fast and economic way
to estimate the sensitivity of a given process on the TGC (or any kind
of `deviation' parameters).
The only thing one has to do is to calculate the 
correlation matrix $\b_{ij}$. This should be contrasted with 
log-likelihood methods where either one has to repeat the calculation
for a large set of TGC parameter values which adequately covers the
parameter space, or to store information on the coefficients
appearing in the \eqn{differentialxs} for each generated `event'.
Of course from the point of view of `unweighted' or real data analysis
these arguments do not apply any more, since now local (event-by-event)
information is any way available. In this case
log-likelihood would be preferable over optimal observables, 
since it is not restricted to the linear terms.

\begin{table}[htb]
\tabcolsep=0pt
\bc
\begin{small}
\begin{tabular}{|c|c|c|c|c|}
\hline 
$\sqrt{s}$ & 161 & 176 & 190 & 205 \\ 
\hline
\hline & &&& \\ 
$e$ 
& 
$  \begin{array}{c}   \mtt{0.456 0.180 0.087}   \\ 
                      \mtt{0.180 0.173 0.008}   \\ 
                      \mtt{0.087 0.008 0.117} \end{array}  $
& 
$  \begin{array}{c}   \mtt{0.587 0.086 0.238}   \\ 
                      \mtt{0.086 0.087 0.014}   \\ 
                      \mtt{0.238 0.014 0.311} \end{array}  $
& 
$  \begin{array}{c}   \mtt{0.973 0.115 0.420}   \\ 
                      \mtt{0.115 0.135 0.024}   \\ 
                      \mtt{0.420 0.024 0.605} \end{array}  $
& 
$  \begin{array}{c}   \mtt{1.400 0.163 0.601}   \\ 
                      \mtt{0.163 0.209 0.036}   \\ 
                      \mtt{0.601 0.036 0.969} \end{array}  $
\\ & & & & \\  \cline{2-5}
& 124.6(6) & 489(2) & 591(2) & 621(2) \\  
\hline \hline & & & &  \\
$\mu$ 
& 
$  \begin{array}{c}   \mtt{0.271 0.013 0.092}   \\ 
                      \mtt{0.013 0.019 0.004}   \\ 
                      \mtt{0.092 0.004 0.093} \end{array}  $
& 
$  \begin{array}{c}   \mtt{0.547 0.033 0.253}   \\ 
                      \mtt{0.033 0.037 0.014}   \\ 
                      \mtt{0.253 0.014 0.313} \end{array}  $
& 
$  \begin{array}{c}   \mtt{0.956 0.063 0.447}   \\ 
                      \mtt{0.063 0.084 0.028}   \\ 
                      \mtt{0.447 0.028 0.620} \end{array}  $
& 
$  \begin{array}{c}   \mtt{1.435 0.111 0.660}   \\ 
                      \mtt{0.111 0.151 0.044}   \\ 
                      \mtt{0.660 0.044 1.036} \end{array}  $
\\ & & & & \\  \cline{2-5}
& 123.1(5) & 476(2) & 567(2) & 585(2) \\  \hline
\end{tabular}
\end{small}
\caption[.]{The correlation matrix for $e$ and $\mu$ channels at LEP2
energies for TGC parameters $\vec{a}=(\awf,\abf,\aw)$. Also shown
are the cross sections as well as their Monte Carlo errors in femtobarns.} 
\label{tab1}
\ec
\end{table}
\par
In \tab{tab1} we present the results for the correlation matrix, $\b_{ij}$,
for centre of mass energies 161, 176, 190 and 205 GeV. 
The total cross sections are also presented.
As is evident from this table, the correlation between the different
$ a$'s are rather strong, which suggests that an analysis taking into
account these correlation is indispensable. As is easily seen from this table,
the correlations are much stronger between $\awf$ and $\aw$ as compared with
$\abf$, which in general gives the smallest contribution.
\par 
An other very interesting result, is that electron and muon channels
exhibit different behaviour depending on the energy. 
The difference is more prominent at the threshold
region, $\sqrt{s}$=161Gev. This is due to the contribution from 
single-resonant $W$-production graph \fig{fig0}, which becomes relatively 
more important at threshold. 
On the other hand at higher energies, the dominance of the 
double-resonant graphs (CC3) restores universality among electrons and muons, 
and the correlation matrix is almost identical for both channels. It is worth
to emphasize that differences in the correlation matrix ($\sim 100\%$) 
among $e$ and $\mu$ channels
are much more important than the differences in the total cross section 
($\sim 1\%$),
reflecting the fact that the graph of \fig{fig0} contributes mainly to the 
shape of the differential cross section rather than to 
its overall normalization.
\begin{table}[htb]
\bc
\begin{small}
\begin{tabular}{|c|c|c|}
\hline 
$\sqrt{s}$ & $e$ & $\mu$  \\ 
\hline\hline & & \\ 
161 
& 
$  \begin{array}{cc} 
        
                            0.559 &\left(\mtt{0.889, 0.419, 0.183}  \right)  \\ 
                            0.125 &\left(\mtt{0.128,-0.613, 0.779}  \right)  \\
                            0.062 &\left(\mtt{0.438,-0.670,-0.599}  \right)  \\
                            \end{array}  $
& 
$  \begin{array}{cc} 
        
                            0.311 &\left(\mtt{0.920, 0.046,-0.390}  \right)  \\ 
                            0.054 &\left(\mtt{0.388, 0.039, 0.921}  \right)  \\
                            0.018 &\left(\mtt{0.058,-0.998,-0.018}  \right)  \\
                            \end{array}  $
\\ & & \\ \hline \hline 
& &  \\  
176 
& 
$  \begin{array}{cc} 
        
                            0.734 &\left(\mtt{0.863, 0.125, 0.489}  \right)  \\ 
                            0.183 &\left(\mtt{0.443, 0.274,-0.853}  \right)  \\
                            0.068 &\left(\mtt{0.241,-0.953,-0.181}  \right)  \\
                            \end{array}  $
& 
$  \begin{array}{cc} 
        
                            0.711 &\left(\mtt{0.841, 0.052, 0.537}  \right)  \\ 
                            0.152 &\left(\mtt{0.535, 0.051,-0.843}  \right)  \\
                            0.035 &\left(\mtt{0.072,-0.997,-0.015}  \right)  \\
                            \end{array}  $
\\ & &  \\  \hline\hline
& &  \\  
190 
& 
$  \begin{array}{cc} 
        
                            1.258 &\left(\mtt{0.835, 0.097, 0.540}  \right)  \\ 
                            0.339 &\left(\mtt{0.517, 0.193,-0.833}  \right)  \\
                            0.116 &\left(\mtt{0.186,-0.976,-0.111}  \right)  \\
                            \end{array}  $
& 
$  \begin{array}{cc} 
        
                            1.269 &\left(\mtt{0.821, 0.057, 0.568} \right)  \\ 
                            0.311 &\left(\mtt{0.565, 0.055,-0.823}  \right)  \\
                            0.080 &\left(\mtt{0.078,-0.997,-0.013}  \right)  \\
                            \end{array}  $
\\ & &  \\  \hline\hline
& &  \\  
205
& 
$  \begin{array}{cc} 
        
                            1.838 &\left(\mtt{0.817, 0.094, 0.569}  \right)  \\ 
                            0.557 &\left(\mtt{0.549, 0.173,-0.817}  \right)  \\
                            0.183 &\left(\mtt{0.175,-0.980,-0.089}  \right)  \\
                            \end{array}  $
& 
$  \begin{array}{cc} 
        
                            1.932 &\left(\mtt{0.802, 0.065, 0.594}  \right)  \\ 
                            0.548 &\left(\mtt{0.589, 0.075,-0.804}  \right)  \\
                            0.141 &\left(\mtt{0.097,-0.995,-0.023}  \right)  \\
                            \end{array}  $
\\ & &  \\  \hline
\end{tabular}
\end{small}
\caption[.]{The eigenvalues and the corresponding eigenvectors of
the correlation matrices given in \tab{tab1}.}
\label{tab2}
\ec
\end{table}
\par
In \tab{tab2} we show the eigenvalues of the correlation matrix, 
as well as the corresponding eigenvectors. These eigenvectors define 
directions in the three-parameter space, which are uncorrelated
and therefore parameter errors can be safely extracted.
As we can see the two dominant eigenvalues correspond to directions 
in the three-parameter space related predominantly to $\awf$ and $\aw$,
whereas the lowest eigenvalue  is always related to $\abf$. 
This $\abf$ suppression is due to the fact that in the double resonant (CC3) graphs
relative cancellations between terms proportional to $\xg$ and $\xZ$ 
take place\footnote{that is cancellations between 
$\gamma$ and $Z$-exchange, $s$-channel graphs.}.
An exception to
this pattern is due to the electron channel at 161 GeV, where the mixing 
is much more important, reflecting the contribution from the graph of \fig{fig0}.
This is easily understood, since in the amplitude corresponding to this graph
the relative cancellations between $\xg$ and $\xZ$ are now destroyed  due to the
the presence of the $t$-channel photon propagator, which means that the 
$t$-channel 
single-resonant graph receives
contributions only from $\xg$ and $\lg$ whereas contributions from 
$\xZ$, $\dZ$ and $\lZ$ are suppressed.
\par 
Finally in \tab{tab3} one-standard-deviation errors are presented
by combining $e$ and $\mu$ in the following
way\footnote{which becomes exact when 
$\langle c_{1,i}/c_0 \rangle\langle c_{1,j}/c_0 \rangle \ll 
\langle \left( c_{1,i}/c_0\right) \left(  c_{1,j}/c_0\right) \rangle$, as
is indeed the case in our calculations.}
\[
\b_{ij}=\b^{(e)}_{ij}{\sigma^{(e)}\over \sigma^{(e)}+\sigma^{(\mu)} }
+\b^{(\mu)}_{ij}{ \sigma^{(\mu)} \over \sigma^{(e)}+\sigma^{(\mu)} }\;\;.
\]
and 
\[ N=4 L ( \sigma^{(e)}+\sigma^{(\mu)} )\;\;. \]
In all our studies, $L$ is taken to be 100 pb$^{-1}$ at 161 GeV
and 500 pb$^{-1}$ at all higher LEP2 energies.
In $1d$-case, corresponding to `one-dimensional log-likelihood fit', 
errors on $\awf$, $\aw$, $\abf$ are shown. For $3d$-cases, the errors 
correspond to the linear combinations defined in \eqn{orthogonal}. 
Nevertheless in \tab{tab3} we kept the notation of $\awf$, $\aw$ and $\abf$
for $3d$-cases, since
the corresponding eigenvector, $\awf^D$ for instance, is mainly composed by 
$\awf$ and so on for the other TGC parameters. 
It is rather evident that correlations, though important, do not
alter dramatically the estimate of the errors based on simple
`one-dimensional' investigations, and therefore $1d$ limits
are still useful.
\begin{table}[htb]
\bc
\begin{tabular}{|c|c|c|c|c|}
\hline 
& $\awf$ \hfill  $\aw$ \hfill  $\abf$ 
& $\awf$ \hfill  $\aw$ \hfill  $\abf$ 
& $\awf$ \hfill  $\aw$ \hfill  $\abf$   
& $\awf$ \hfill  $\aw$ \hfill  $\abf$ \\  
$\sqrt{s}$ & 161 & 176 & 190 & 205 
 \\ \hline\hline
$1d$ 
& $ \mtt{0.16  0.31  0.32} $
& $ \mtt{0.030  0.041  0.091} $
& $ \mtt{0.021  0.026  0.062} $
& $ \mtt{0.017  0.020  0.048} $
\\  \hline \hline
$3d$ 
& $ \mtt{0.15  0.32  0.44} $ 
& $ \mtt{0.027  0.056  0.098} $ 
& $ \mtt{0.018  0.036  0.066} $ 
& $ \mtt{0.015  0.027  0.050} $ 
\\  \hline \hline
$3d^\prime$ 
& $ \mtt{0.18  0.37  0.50} $ 
& $ \mtt{0.031  0.064  0.110} $ 
& $ \mtt{0.021  0.041  0.073} $ 
& $ \mtt{0.017  0.030  0.055} $ 
\\   \hline
\end{tabular}
\caption[.]{One standard deviation errors on TGC parameters.
$1d$ means that the correlation matrix is assumed diagonal where
for $3d$ and $3d^\prime$ all correlations have been taken into account.
Finally in $3d^\prime$ `resolution' effects are also included as described 
in the text.}
\label{tab3}
\ec
\end{table}


\par
In real life, detector resolution effects and reconstruction algorithm, 
tend to distort the `observed' distributions
as compared with the `theoretical' predictions. This can in principle,
mimic the effects of TGC, and therefore should be studied on a 
generator level. This study requires a good knowledge of the specific
detector and should be carried out in detail by our expiremental colleagues. 
Nevertheless, we can estimate the order of magnitude of 
these effects based on the following very general
assumption, that the distorted-`observed' distribution $p_{eff}$
is expressed as a convolution integral given by: 
\bq 
p_{eff}(\Omega_{meas},\vec{a})=\int p(\Omega_{true},\vec{a}) 
\;\;S(\Omega_{true},\Omega_{meas})\;\; d\Omega_{true}
\label{convolution}
\eq 
where $p(\Omega_{true},\vec{a})$ is the theoretically 
predicted distribution and $S(\Omega_{true},\Omega_{meas})$
is a `resolution' function, giving the probability that an event
`measured' at phase space point $\Omega_{meas}$ is coming from 
a `true' event at a phase space point $\Omega_{true}$.
\par In order to simulate in a simple way the `resolution' function we have
implemented the following algorithm:
\begin{itemize}
\item To each generated event we calculate the nine kinematical variables, 
in the lab frame, $E_i,\theta_i,\phi_i$, $i=1,2,3$,
which correspond to the `visible' particles, namely the lepton $\ell$ and
the two jets $q,\bqp$. 
\item To each of the above variables we assign a new one, 
\[ \theta\to\theta^\prime,\;\;\phi\to\phi^\prime\;\;\mbox{and}\;\;
E\to E^\prime
\]
following the rule that primed
variables are normally distributed around
the original ones\footnote{ with the exception
of jet energies, where $\langle E^\prime \rangle = 0.85\;E$, 
in order to take into account energy losses in jet reconstruction.}
with variances given by:
\bqa
&
\sigma_{E_{\ell}}=0.03\;E_{\ell},\;
\sigma_{\theta_{\ell}}=0.0002,\;
\sigma_{\phi_{\ell}}=0.0002,\; & \nn \\
&
\sigma_{E_{jet}} = 0.15\;E_{jet},\;
\sigma_{\theta_{jet}}=0.03,\;
\sigma_{\phi_{jet}}=0.04\;. \nn &
\eqa
\item Then, this artificially generated phase space point is subjected 
to a `kinematical constraint fit' algorithm, by including a `missing energy'
(neutrino) four-momentum in such a away that
\[ \sum_{i=1}^{4} E_i =2\;E_{beam}\;\;\;\mbox{and}\;\;\;
\sum_{i=1}^{4} \vec{p}_i =0
\] 
This is achieved by a local (event by event) rescaling of the energies
of the two jets, which are expected to be the worst measured quantities.
This rescaling is defined by the minimization of the following
$\chi^2$ function:
\[
\chi^2=\sum_{jets} { \left( E_i^\prime - E_i\right)^2\over \sigma^2_i\;}.
\] 
Note that, although the original phase space point is generated taking into account
initial state radiation, in the `reconstructed' one 
we have artificially neglected it and the total energy is normalized to the overall
constant beam energy. 
\item Finally the usual cuts of \eqn{cuts} are applied, including now a special
cut on the missing $\notp_T$ vector, $\notp_T\geq 15\gev $. 
This supplementary cut is applied in order to minimize the background
coming from the process
$e^+e^-\to e^+e^-\; q \bar{q}$, where the $e^+$ is assumed to be lost 
into the beam, which leads to the same final state $e^- + 2 \mbox{jets}$.
Using ERATO we have checked that the reconstructed $\notp_T$ in the
background process is well below 15 GeV, 
whereas the $\notp_T$ distribution of our `signal' processes 
$e^+e^-\to \ell\bar{\nu}_{\ell}\; q \bar{q}$ has a negligible tail below 15 GeV.
\end{itemize}

Using now the kinematical information from the above algorithm we can 
calculate the new correlation matrix, $\tilde{\b}_{ij}$, using \eqn{convolution},
and from it the new information matrix given by
\[
I^{new}=I\cdot\tilde{I}^{-1}\cdot I 
\]

The 1sd-erros on TGC parameters calculated through $I^{new}$ 
are given in \tab{tab3}, under the $3d^\prime$-case. 
Comparing with the $3d$ case, we see that 
the effect is of the order of 10-20\% on the parameter errors, which 
is at an acceptable level. We have checked that reasonable input values
for the variances $\sigma_E$, $\sigma_\theta$ and $\sigma_\phi$ 
give more or less the same results. The same is still true if we 
replace the polar representation of the observed momenta by the cartesian one.
This means that resolution effects
are unlikely to destroy the measurement of TGC. Moreover it is worth to
emphasize that `unfolding' techniques can also be used in order to
minimize this effect, relying on a good knowledge of the
detector as well as of the physics included in the event generator programme. 

\par We conclude by summarizing the results of our study:
\ben
\item Sensitivities on TGC at LEP2 are of the level
      of 0.15-0.5 (1sd) at 161 GeV, making the threshold  phase of LEP2
      the best up to date world measurement of TGC. Higher energies
      will drastically improve these limits, almost 
      by an order of magnitude, reaching the level of 0.02.
      Taking into account information from other channels, such 
      as the four-jet one, further improvement is expected.
\item The electron channel due to single-resonant $W$ contributions 
      is rather important at 161 GeV and special
      studies are needed on the experimental side, in order to maximize
      selection efficiencies in this channel. Finite-width, ISR and 
      Coulomb corrections are also indispensable at 161 GeV.
\item Resolution effects are unlikely to drastically alter this picture, 
      leading to a mild loosening of the sensitivity limits by a 10-20\% 
      factor.
\een

\begin{figure}[h]
\begin{center}
\mbox{\psfig{file=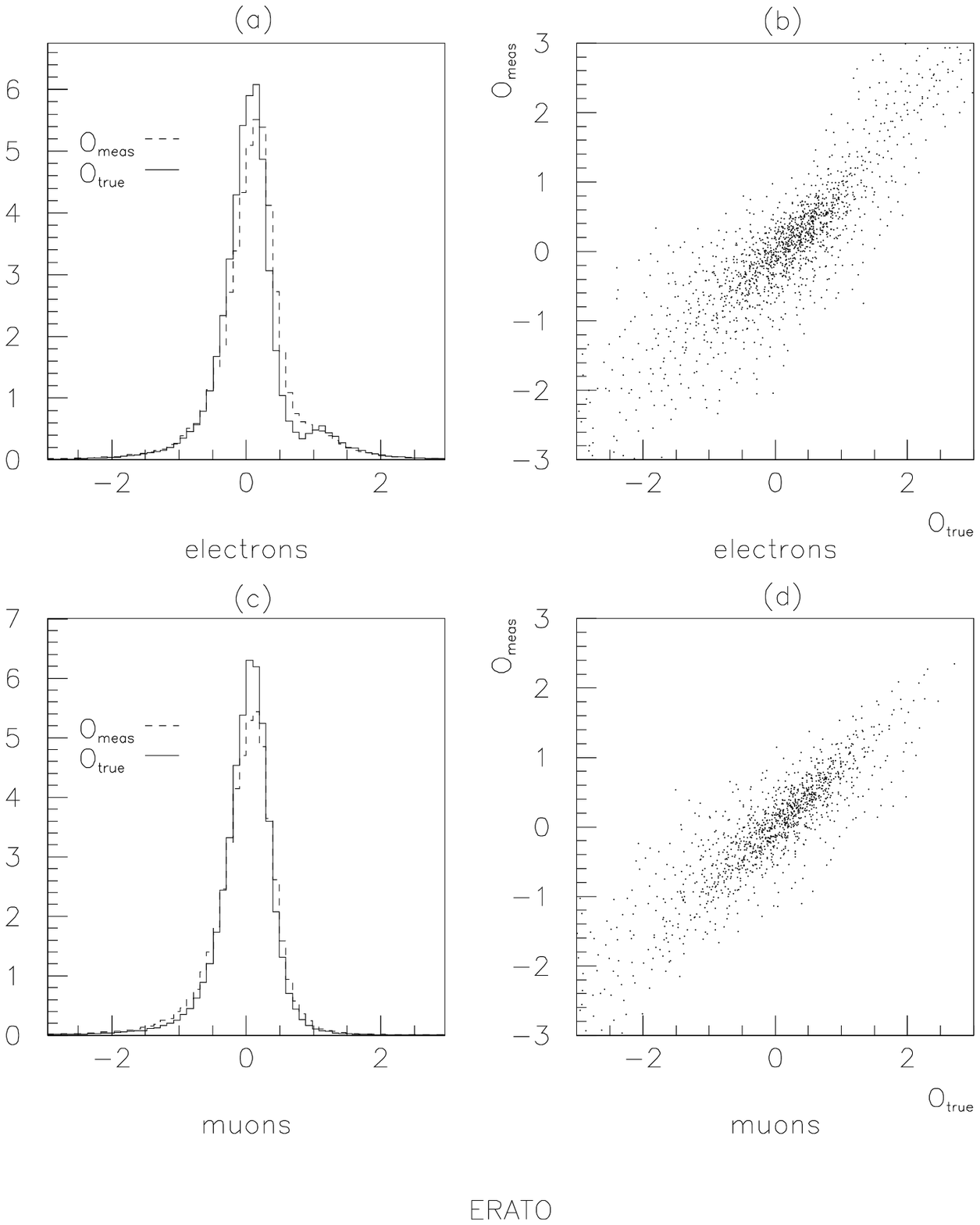,width=15cm,height=15cm}}
\caption[.]{The distribution of the optimal observable corresponding to
the parameter $\awf$ at 161 GeV, for electron (a) and muon (c) channels. Also
shown are the scatter plots (b) and (d).}
\label{fig1}
\end{center}
\end{figure}

As a postscript of our study we present in \fig{fig1} the 
distributions of the optimal observable $\O_{\awf}$ at 161GeV. The solid line
corresponds to the ideal case, where the measured distribution is assumed
to be identical with the predicted one, whereas the dashed line 
is calculated by the convolution integral \eqn{convolution}, based  
on our naive `resolution' model. Also shown are the two-dimensional
scatter plots, between the generated and `measured' values of $\O_{\awf}$.
Although our `resolution' model is very simplistic, it suggests that 
optimal observables can be used in the
experimental analysis of four-fermion data. Optimal observables
are  the `phase-space' variables exhibiting the maximum sensitivity 
on TGC and therefore they effectively reduce the number of independent 
phase-space variables from eight to three, which is rather appreciable
from the experimental point of view. 
On the other hand
their sensitivity on `resolution' effects seems to be rather mild.
Nevertheless their usefulness has to be verified by a detailed analysis at the 
experimental level.
      
\vspace*{1cm}
\noindent {\bf Acknowledgements} \\[12pt]
It is pleasure to thank H. Phillips, R.Sekulin and S.Tzamarias for 
making the simulation values of the experimental resolutions available to 
the author. 
This work is supported in part by the EU grant CHRX-CT93-0319.

\end{document}